\documentclass[twocolumn,epj]{svjour}

\usepackage{hyperref}
\usepackage{graphicx,colordvi}
\usepackage{amsmath,amssymb,slashed,cite}
\usepackage{booktabs,tabulary}
\usepackage{dsfont}
\usepackage{color}
\usepackage{epstopdf}

\newcommand{\diff}{\text{d}}

\newcommand{\GeV}{\,\text{GeV}}
\renewcommand{\Im}{\text{Im}\,}

\newcommand{\fm}{\,\text{fm}}
\newcommand{\mpi}{M_\pi} 
\newcommand{\mpii}{M_{\pi^0}}
\newcommand{\mN}{m_N}
\newcommand{\mpp}{m_p}
\newcommand{\mn}{m_n}
\newcommand{\Fpi}{F_\pi}
\newcommand{\ga}{g_A}
\newcommand{\Order}{\mathcal{O}}
\newcommand{\eps}{\epsilon}
\newcommand{\tpi}{t_\pi}
\newcommand{\tN}{t_N}
\newcommand{\beq}{\begin{equation}}
\newcommand{\eeq}{\end{equation}}
\newcommand{\bsp}{\begin{sloppypar}}
\newcommand{\esp}{\end{sloppypar}}

\newcommand{\toright}[1]{\hspace*{\fill}{\footnotesize{#1}}}

\allowdisplaybreaks[1]

\begin{document}

\title{\toright{\textnormal{INT-PUB-16-031}}\\[0.1cm]On the $\boldsymbol{\pi\pi}$ continuum in the nucleon form factors and the proton radius puzzle}
\titlerunning{On the $\pi\pi$ continuum in the nucleon form factors and the proton radius puzzle}
\author{M.\ Hoferichter\inst{1}
	\and
        B.\ Kubis\inst{2}
        \and
        J.\ Ruiz de Elvira\inst{2}
        \and
        H.-W.\ Hammer\inst{3,4}
        \and
        U.-G.\ Mei{\ss}ner\inst{2,5}
}
\institute{Institute for Nuclear Theory, University of Washington, Seattle, WA 98195-1550, USA
\and
Helmholtz--Institut f\"ur Strahlen- und Kernphysik (Theorie) and
   Bethe Center for Theoretical Physics, Universit\"at Bonn, D--53115 Bonn, Germany
\and   
Institut f\"ur Kernphysik, Technische Universit\"at Darmstadt, D--64289 Darmstadt, Germany
\and
ExtreMe Matter Institute EMMI, GSI Helmholtzzentrum f\"ur Schwerionenforschung GmbH, D--64291 Darmstadt, Germany   
\and
Institut f\"ur Kernphysik, Institute for Advanced Simulation, and 
   J\"ulich Center for Hadron Physics, Forschungszentrum J\"ulich, D--52425  J\"ulich, Germany
}

\date{}

\abstract{We present an improved determination of the $\pi\pi$ continuum contribution to the isovector spectral functions of the nucleon electromagnetic form factors. Our analysis includes
the most up-to-date results for the $\pi\pi\to\bar N N$ partial waves extracted from Roy--Steiner equations, consistent input for the
pion vector form factor, and a thorough discussion of isospin-violating effects and uncertainty estimates. As an application, we consider the
$\pi\pi$ contribution to the isovector electric and magnetic radii by means of sum rules, which, in combination with the accurately known neutron electric radius, are found to slightly prefer a small proton charge radius.
}

\maketitle

\section{Introduction}
\bsp
One of the most startling discoveries in low-energy hadron physics over the last years concerns the size of the proton,
with spectroscopy measurements in muonic atoms~\cite{Pohl:2010zza,Antognini:1900ns,Pohl:2016xxx} 
revealing that the electric charge radius of the proton might be significantly smaller than previously thought,
with the average of extractions from hydrogen spectroscopy, $r_E^p=0.8758(77)\fm$~\cite{Mohr:2012tt}, $4.5\sigma$ larger than the result $r_E^p=0.84087(39)\fm$ from muonic hydrogen.
A third, independent method to determine the proton radius relies on low-energy electron--proton scattering experiments, extracting radii from the slope of the form factors.
It is by combining hydrogen spectroscopy and scattering data that CODATA~\cite{Mohr:2012tt} quoted $r_E^p=0.8775(51)\fm$, a $7\sigma$ deviation from muonic spectroscopy. 
For a more detailed discussion of this proton radius puzzle, including potential explanations by physics beyond the Standard Model and pathways towards its resolution by new experiments, we refer to~\cite{Pohl:2013yb,Carlson:2015jba}.

Evidently, even besides the possibility of error on the experimental side, exotic explanations for the proton-radius discrepancy such as~\cite{Brax:2010gp,Barger:2010aj,TuckerSmith:2010ra,Batell:2011qq,Barger:2011mt,Carlson:2012pc,Wang:2013fma,Onofrio:2013fea,Karshenboim:2014tka,Carlson:2015poa,Liu:2016qwd}
can only be seriously considered if the hadronic physics in each system is thoroughly understood.
Here, we address aspects of the radius extraction from scattering data, the systematics of which have been discussed vigorously in the literature recently~\cite{Bernauer:2010wm,Hill:2010yb,Lorenz:2012tm,Bernauer:2013tpr,Lorenz:2014vha,Epstein:2014zua,Lorenz:2014yda,Lee:2015jqa,Griffioen:2015hta,Hill:2016gdf,Bernauer:2016ziz}.
Irrespective of the challenges involved in data selection and radiative corrections, the extraction can be stabilized by respecting the analytic structure of the electromagnetic form factors~\cite{Frazer:1960zzb}, e.g., by employing a conformal expansion.
Going one step further, a dispersion theoretical analysis of nucleon form factors~\cite{Hohler:1976ax,Hohler:1974ht,Hohler:1974eq,Mergell:1995bf,Hammer:2003ai,Belushkin:2005ds,Belushkin:2006qa,Lorenz:2012tm} approaches the extraction by expressing the spectral functions in terms of the lowest intermediate states as well as effective couplings, and reconstructing the full form factor by means of a dispersion relation supplemented by superconvergence relations that follow from perturbative QCD~\cite{Lepage:1980fj}. 
In both cases the most important contribution is generated by the $\pi\pi$ continuum, by definition in a dispersive analysis, but also in a conformal expansion, where the explicit consideration of 
$\pi\pi$ intermediate states allows one to increase the threshold in the conformal mapping.
It is worth noting that dispersive analyses of the nucleon form factor
data including the $\pi\pi$ continuum in the spectral function
consistently gave a small proton charge radius even before the emergence
of the radius puzzle~\cite{Hohler:1976ax,Hohler:1974ht,Mergell:1995bf,Hammer:2003ai,Belushkin:2006qa}.
In principle, the $\pi\pi$ continuum follows unambiguously from the partial waves for
$\pi\pi\to\bar N N$ analytically continued into the pseudophysical region as well as the pion vector form factor, but the last comprehensive analysis of the $t$-channel partial waves dates back to~\cite{Hoehler:1983}, based on the Karlsruhe--Helsinki partial-wave analysis~\cite{Koch:1980ay}. 
Apart from outdated data input, the resulting spectral function has been criticized due to its lack of a thorough analysis of systematic uncertainties and consideration of isospin-violating corrections~\cite{Lee:2015jqa}. These shortcomings will be remedied in the present paper. 

Due to crossing symmetry, the partial waves for $\pi\pi\to\bar N N$ occur naturally in a dispersive analysis of pion--nucleon ($\pi N$) scattering based on Roy--Steiner (RS) equations~\cite{Ditsche:2012fv,Hoferichter:2012wf,Hoferichter:2015dsa,Hoferichter:2015tha,Hoferichter:2015hva,Hoferichter:2016ocj,Siemens:2016jwj},
in fact, in the same way as for nucleon form factors, it is the pseudophysical region $t\geq 4\mpi^2$ that dominates the dispersion integrals. In addition to the error estimates 
for the partial waves, we construct a description of the pion vector form factor fit to the latest experimental results from BaBar~\cite{Aubert:2009ad}, KLOE~\cite{Babusci:2012rp}, and BESIII~\cite{Ablikim:2015orh} and consistent with the input for the $\pi\pi$ $P$-wave phase shifts used in the solution of the RS system.
This procedure ensures that the resulting spectral function fulfills all available constraints from modern low-energy data, including the precision measurements of $\pi N$ dynamics extracted from
pionic atoms~\cite{Strauch:2010vu,Hennebach:2014lsa,Gotta:2008zza,Baru:2010xn,Baru:2011bw}.
In addition to their relevance in stabilizing form-factor fits, the resulting spectral functions also find applications in improving the
nucleon-pole-term contribution to bremsstrahlung of a $\rho$ in proton--nucleus scattering~\cite{deNiverville:2016rqh} and in the 
transverse charge densities of the nucleon~\cite{Miller:2011du}.   

After reviewing the necessary formalism in Sect.~\ref{sec:formalism}, we discuss the role of isospin-violating effects in detail in Sect.~\ref{sec:isosin}.
The results for the isovector spectral functions for the electric and magnetic nucleon form factors are provided in Sect.~\ref{sec:spectral}, with
applications to sum rules for form-factor normalizations and radii considered in Sect.~\ref{sec:sum_rules}. 
Our conclusions are summarized in Sect.~\ref{sec:conclusions}.

\section{Formalism}
\label{sec:formalism}

\subsection{Nucleon form factors}

\begin{figure}[t!]
\centering
\includegraphics[width=0.35\linewidth]{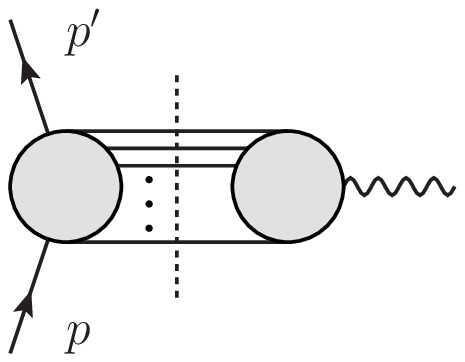} \
\raisebox{1cm}{$=$} \
\includegraphics[width=0.35\linewidth]{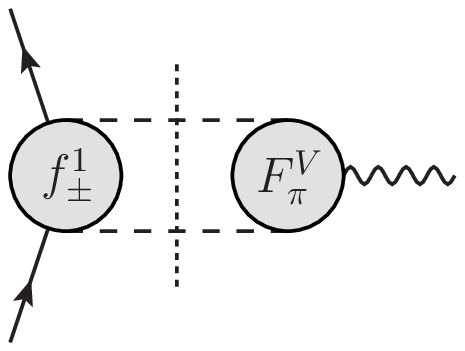} \
\raisebox{1cm}{$\ + \ \cdots$} 
\caption{Unitarity relation for nucleon form factors. Solid, dashed, and wiggly lines denote nucleons, pions, and the external photon, respectively, while the short-dashed lines indicate the cutting of particle propagators. In the isovector channel, the sum over all intermediate states as represented by the first diagram starts with $\pi\pi$, whose contribution is fixed in terms of the $\pi\pi\to\bar N N$ $P$-waves $f^1_\pm$ and the pion vector form factor $F_\pi^V$, see main text.}
\label{fig:unitarity}
\end{figure}

Based on Lorentz, gauge, and parity invariance the matrix element of the electromagnetic current 
\beq
j^\mu_\text{em}=\bar q \mathcal{Q} \gamma^\mu q,
\eeq
with light quarks $q=(u,d,s)^T$ and charges $\mathcal{Q}=\text{diag}(2,-1,-1)/3$ (in units of the elementary charge $e$), between nucleon states admits the decomposition
\beq
\langle N(p')|j^\mu_\text{em}|N(p)\rangle=\bar u(p')\bigg[F_1^N(t)\gamma^\mu+\frac{i \sigma^{\mu\nu}q_\nu}{2\mN} F^N_2(t)\bigg]u(p),
\eeq
where $\mN$ refers to the nucleon mass ($N=p,n$), $F_1^N$ ($F_2^N$) to the Dirac (Pauli) form factor, and
$q=p'-p$, $t=q^2$. The form factors are normalized according to
\beq
F_1^p(0)=1,\qquad F_1^n(0)=0,\qquad F_2^N(0)=\kappa_N,
\eeq
with anomalous magnetic moments $\kappa_p=1.792847356(23)$ and $\kappa_n=-1.91304272(45)$~\cite{Agashe:2014kda}.
We further define isoscalar and isovector combinations as
\begin{align}
F_i^s(t)&=\frac{1}{2}\big(F_i^p(t)+F_i^n(t)\big),\notag\\
F_i^v(t)&=\frac{1}{2}\big(F_i^p(t)-F_i^n(t)\big),
\end{align}
and use the Sachs form factors
\begin{align}
G_E^N(t)&=F_1^N(t)+\frac{t}{4\mN^2}F_2^N(t),\notag\\
G_M^N(t)&=F_1^N(t)+F_2^N(t).
\end{align}
The electric and magnetic radii are defined as $r=\sqrt{\langle r^2\rangle}$ with
\begin{align}
\label{radii}
 \langle r^2_E\rangle^p&=6\bigg[\frac{\diff}{\diff t}G_E^p(t)\bigg]_{t=0},\notag\\
 \langle r^2_M\rangle^p&=\frac{6}{1+\kappa_p}\bigg[\frac{\diff}{\diff t}G_M^p(t)\bigg]_{t=0},\notag\\
 \langle r^2_E\rangle^n&=6\bigg[\frac{\diff}{\diff t}G_E^n(t)\bigg]_{t=0},\notag\\
 \langle r^2_M\rangle^n&=\frac{6}{\kappa_n}\bigg[\frac{\diff}{\diff t}G_M^n(t)\bigg]_{t=0}.
\end{align}
The form factors fulfill dispersion relations of the form
\beq
\label{disp}
F(t)=\frac{1}{\pi}\int\limits_{t_0}^\infty\diff t'\,\frac{\Im F(t')}{t'-t-i\eps},
\eeq
where the threshold depends on the allowed intermediate states. In the isospin limit, $t_0=4(9)\mpi^2$ for the isovector (isoscalar) combinations. This strict separation fails once isospin violation is permitted, as detailed in Sect.~\ref{sec:rho_omega}. In particular, in the presence of isospin-violating corrections all contributions will be classified with respect to the nucleon states in which they
occur, i.e., isoscalar (isovector) terms are defined to contribute to the sum (difference) of proton and neutron. 
In the isovector channel the leading contribution to $\Im F$ is generated by $\pi\pi$ intermediate states, see Fig.~\ref{fig:unitarity}. By cutting the pion propagators, it is clear that the corresponding imaginary part is related to the on-shell amplitudes for $\pi\pi\to\bar N N$ and $\pi\pi\to\gamma^*$, a relation to be made more precise in the following subsections. 

\subsection{Pion--nucleon scattering}

In the isospin limit, the amplitude for $\pi N$ scattering
\beq
\pi^a(q)+N(p)\to\pi^b(q')+N(p'),
\eeq
with pion isospin labels $a$, $b$ and Mandelstam variables
\beq
s=(p+q)^2,\qquad t=(p'-p)^2,\qquad u=(p-q')^2,
\eeq
reduces to two scalar functions
\begin{align}
T^{ba}(s,t)&=\delta^{ba}T^+(s,t)+\frac{1}{2}[\tau^b,\tau^a]T^-(s,t),\\
T^I(s,t)&=\bar{u}(p')\bigg\{A^I(s,t)+\frac{1}{2}(\slashed q + \slashed{q'})B^I(s,t)\bigg\}u(p),\notag
\end{align}
where $\tau^a$ are isospin Pauli matrices and $I=\pm$ refers to isoscalar/isovector amplitudes.

For the unitarity relation of the nucleon form factors we need 
the $t$-channel partial waves~\cite{Frazer:1960zza}
\begin{align}
\label{tprojform}
f^J_+(t)&=-\frac{1}{4\pi}\int\limits^1_0\diff z_t\;P_J(z_t)\bigg\{\frac{p_t^2}{(p_tq_t)^J}A^I-\frac{\mN z_t}{(p_tq_t)^{J-1}} B^I\bigg\},\notag\\
f^J_-(t)&=\frac{1}{4\pi}\frac{\sqrt{J(J+1)}}{2J+1}\frac{1}{(p_tq_t)^{J-1}}\notag\\
&\quad\times\int\limits^1_0\diff z_t\Big[P_{J-1}(z_t)-P_{J+1}(z_t)\Big]B^I,
\end{align}
with $t$-channel scattering angle 
\beq
z_t=\frac{s-u}{4p_tq_t},
\eeq
momenta
\begin{align}
p_t&=\frac{1}{2}\sqrt{t-\tN}, &q_t&=\frac{1}{2}\sqrt{t-\tpi},\notag\\
\tN&=4\mN^2,&\tpi&=4\mpi^2,
\end{align}
and angular momentum $J$ ($I=\pm$ corresponds to even/odd $J$).
The arguments of the invariant amplitudes in~\eqref{tprojform} are to be expressed as $s=s(t,z_t)$.
Due to the quantum numbers of the external current, the unitarity relation projects out the $P$-waves $f^1_\pm$. 

Finally, we need the subthreshold expansion of the $\pi N$ amplitudes, which proceeds in terms of
$\nu=(s-u)/(4\mN)$ and $t$ around $\nu=t=0$ 
\begin{align}
\label{subthr}
\bar A^\pm(\nu,t)&=\begin{pmatrix}1\\\nu\end{pmatrix}
\sum_{n,m=0}^\infty a_{mn}^\pm\nu^{2m}t^n,\notag\\
\bar B^\pm(\nu,t)&=\begin{pmatrix}\nu\\1\end{pmatrix}
\sum_{n,m=0}^\infty b_{mn}^\pm\nu^{2m}t^n,
\end{align}
where the upper/lower entry corresponds to $I=\pm$, 
and the 
Born-term-subtracted amplitudes are defined as
\beq
\bar X^\pm(\nu,t)=X^\pm(\nu,t)-X^\pm_\text{pv}(\nu,t),\quad
X\in\{A,B\},
\eeq
with
\begin{align}
\label{Born_terms}
B^\pm_\text{pv}(\nu,t)&=g^2\bigg(\frac{1}{\mN^2-s}\mp\frac{1}{\mN^2-u}\bigg)
-\frac{g^2}{2\mN^2}\begin{pmatrix}0\\1\end{pmatrix},\notag\\
A^\pm_\text{pv}(\nu,t)&=\frac{g^2}{\mN}\begin{pmatrix}1\\ 0\end{pmatrix}.
\end{align}
$g$ denotes the $\pi N$ coupling constant, to be identified later with $g_c$ for the charged-pion vertex. Numerically, we will use
$g_c^2/(4\pi)=13.7(0.2)$~\cite{Baru:2011bw}, in line with the most recent determination from nucleon--nucleon scattering~\cite{Perez:2016aol}.

\subsection{Pion vector form factor}

The electromagnetic form factor of the pion is defined as
\beq
\langle \pi^+(p')|j^\mu_\text{em}|\pi^+(p)\rangle =(p+p')^\mu F_\pi^V(t).
\eeq
The $\pi\pi$ intermediate states produce the unitarity relation
\beq
\label{unitarity_FpiV}
\Im F_\pi^V(t)=\sin\delta^1_1(t) e^{-i\delta^1_1(t)} F_\pi^V(t) \theta\big(t-\tpi\big),
\eeq
with the $\pi\pi$ $P$-wave phase shift $\delta^1_1$. Eq.~\eqref{unitarity_FpiV} reflects Watson's final-state theorem~\cite{Watson:1954uc},
which states that the phase of $F_\pi^V$ has to coincide with the $\pi\pi$ scattering phase shift (up to multiple integers of $\pi$),
to ensure that the imaginary part on the left-hand side of the equation stays real. 
In fact, neglecting higher intermediate states unitarity determines $F_\pi^V(t)$ up to a polynomial $P(t)$ in terms of the Omn\`es factor $\Omega^1_1(t)$~\cite{Omnes:1958hv}
\beq
\label{FpiOmnes}
 F_\pi^V(t)=P(t)\Omega^1_1(t)
 =P(t)\exp\Bigg\{\frac{t}{\pi}\int\limits_{\tpi}^{\infty}\diff t'\frac{\delta^1_1(t')}{t'(t'-t)}\Bigg\}.
\eeq
In practice, the representation~\eqref{FpiOmnes} indeed provides a very efficient parameterization of the experimental data, up to the distortions due to $\rho$--$\omega$ mixing.
To include this isospin-violating effect, we use
\beq
\label{FpiV}
F_\pi^V(t) = \bigg(1+\alpha t+ \frac{\eps\,t}{M_\omega^2-iM_\omega\Gamma_\omega-t} \bigg)\Omega_1^1(t), 
\eeq
with $\omega$ mass $M_\omega$ and width $\Gamma_\omega$. $\alpha$ and $\eps$ are fit to~\cite{Aubert:2009ad,Babusci:2012rp,Ablikim:2015orh} below $\sqrt{t}=1\GeV$, using
the same $\pi\pi$ phase shifts as in the RS analysis~\cite{Hoferichter:2015hva}, determined from Roy and Roy-like equations by the Bern~\cite{Caprini:2011ky,Colangelo:2001df}  
and the Madrid--Cracow group~\cite{GarciaMartin:2011cn}. We also use a variant of the Bernese phase shift that includes effects from $\rho'$ and $\rho''$ in an elastic approximation~\cite{Schneider:2012ez}.
The sensitivity to the three different phase shifts and data sets will be part of the uncertainty estimate for the final spectral functions.

\subsection{Unitarity relation}

Taking everything together, the unitarity relations for the nucleon form factors become~\cite{Frazer:1960zzb}
\begin{align}
\label{unitarity_relation}
 \Im G_E^v(t)&=\frac{q_t^3}{\mN\sqrt{t}}\big(F_\pi^V(t)\big)^*f_+^1(t)\theta\big(t-\tpi\big),\notag\\
  \Im G_M^v(t)&=\frac{q_t^3}{\sqrt{2t}}\big(F_\pi^V(t)\big)^*f_-^1(t)\theta\big(t-\tpi\big).
\end{align}
Watson's theorem again ensures that the left-hand side of the equations stays real, as long as the same $\pi\pi$ phase shift is used in the calculation of the pion form factor
and the $\pi\pi\to\bar N N$ partial waves, which is the reason why we consider the same three variants of $\delta^1_1$ in the data fits for $F_\pi^V$ as in the RS analysis of~\cite{Hoferichter:2015hva}. 
The full consistency among all ingredients entering the unitarity relation that is achieved in this way constitutes a key improvement over previous calculations.

In this context we comment on the range of validity of the $2\pi$ approximation. Strictly speaking, the $4\pi$ threshold opens at $\sqrt{t}=4\mpi=0.56\GeV$, but it is well known phenomenologically that the $4\pi$ contribution is completely negligible below the $\omega\pi$ threshold at $\sqrt{t}=0.92\GeV$~\cite{Eidelman:2003uh} (see also~\cite{Hanhart:2012wi}), and only becomes sizable once the $\rho'$, $\rho''$ resonances are excited. For this reason, we restricted the form factor fits to the energy region below $1\GeV$. We will show results up to the two-nucleon threshold, but due to the neglect of $4\pi$ intermediate states, the final nucleon spectral function will be less reliable beyond $1\GeV$, where such effects might become important. However, while a complete calculation would require the solution of a coupled-channel system of $2\pi$ and $4\pi$ for both $F_\pi^V$ and $f^1_\pm$, the variant of the $\pi\pi$ phase shift from~\cite{Schneider:2012ez} is constructed in such a way that a single-channel Omn\`es representation reproduces $\rho'$ and $\rho''$ effects as manifest in $\tau\to\pi\pi\nu_\tau$ data~\cite{Fujikawa:2008ma}.
The deviation of the corresponding result, taken to be our central solution, from the variants without $\rho'$, $\rho''$ admixture should therefore provide a realistic estimate of the potential impact of $4\pi$ intermediate states.

\section{Isospin-violating corrections}
\label{sec:isosin}

The isospin conventions in the RS analysis of~\cite{Hoferichter:2015hva} are chosen in such a way that the amplitudes in $I=\pm$ basis are determined based on the charged-pion--proton channels $\pi^\pm p\to\pi^\pm p$ with virtual photons removed, which is already very close to the amplitudes needed for the nucleon spectral functions. 
To identify additional corrections, we turn to the explicit form of the Muskhelishvili--Omn\`es~\cite{Muskhelishvili:1953,Omnes:1958hv} representation for $f^1_\pm(t)$ used in~\cite{Hoferichter:2015hva}
\begin{align}
\label{MO}
\Gamma^1(t)&=\Delta_\Gamma^1(t)+\frac{p_t^2}{12\pi \mN}\Big\{a_{00}^-\big(1-t\,\dot\Omega^1_1(0)\big)+a_{01}^-t\Big\}\Omega^1_1(t)\notag\\
&+\frac{t^2(t-\tN)\Omega^1_1(t)}{\pi}\int\limits_{\tpi}^{\tN}\diff t'\frac{\Delta_\Gamma^1(t')\sin\delta^1_1(t')}{t'^2(t'-\tN)(t'-t)|\Omega^1_1(t')|},\notag\\
f^1_-(t)&=\Delta^1_-(t)+\frac{t^2\Omega^1_1(t)}{\pi}\int\limits_{\tpi}^{\tN}\diff t'\frac{\Delta^1_-(t')\sin\delta^1_1(t')}{t'^2(t'-t)|\Omega_1(t')|}\notag\\
&+\frac{\sqrt{2}}{12\pi}\bigg\{\bigg(b_{00}^--\frac{g^2}{2\mN^2}\bigg)\big(1-t\,\dot\Omega^1_1(0)\big)+b_{01}^-t\bigg\}\Omega^1_1(t),\notag\\
\Gamma^1(t)&=\frac{\mN}{\sqrt{2}}f^1_-(t)-f^1_+(t),\notag\\
\Delta_\Gamma^1(t)&=\frac{\mN}{\sqrt{2}}\Delta^1_-(t)-\Delta^1_+(t),
\end{align}
where in this context $\Omega^1_1(t)$ refers to an Omn\`es function with finite cutoff $\tN$ in the integral, $\dot\Omega^1_1(0)$ its derivative at $t=0$, and 
the inhomogeneities $\Delta^1_\pm(t)$ comprise contributions from Born terms as well as the crossed-channel $\pi N$ partial waves.

The first additional isospin-violating corrections would thus be expected from the proton--neutron mass difference in the Born terms and neutral-pion contributions to
the subthreshold parameters (see~\eqref{subthr} for their precise definition). Moreover, similarly to the pion vector form factor, there will be a contribution from $\rho$--$\omega$ mixing in the unitarity relation. In the following, we will consider each class of these potential corrections in detail.

\subsection{Born terms}

In chiral perturbation theory (ChPT) the $\pi^\pm p\to\pi^\pm p$ Born terms with physical intermediate states take the form
\begin{align}
\label{AB_physical}
A_{\pi^- p\to\pi^-p}&=\frac{\tilde g^2(\mpp+\mn)}{2\Fpi^2}\frac{s-\mpp^2}{s-\mn^2},\notag\\
A_{\pi^+ p\to\pi^+ p}&=\frac{\tilde g^2(\mpp+\mn)}{2\Fpi^2}\frac{u-\mpp^2}{u-\mn^2},\notag\\
B_{\pi^-p\to\pi^-p}&=-\frac{\tilde g^2}{2\Fpi^2}\bigg(1+\frac{(\mpp+\mn)^2}{s-\mn^2}\bigg),\notag\\
B_{\pi^+p\to\pi^+ p}&=\frac{\tilde g^2}{2\Fpi^2}\bigg(1+\frac{(\mpp+\mn)^2}{u-\mn^2}\bigg),
\end{align}
where $\tilde g$ denotes the chiral-limit value of the axial charge $\ga$.
In this way, we can identify the $\pi N$ coupling constant $g_c$ for the charged-pion vertex via the residue of the $B^\pm$-amplitudes as
\beq
g_c=\frac{\tilde g (\mpp+\mn)}{2\Fpi},
\eeq
which in the isospin limit (and ignoring higher-order corrections) indeed reduces to the Goldberger--Treiman relation $g_c=\ga \mN/\Fpi$.
Formulated in terms of the isospin basis, the relevant Born amplitudes therefore become
\begin{align}
\label{AB_isospin}
 A^+&=\frac{2g_c^2}{\mpp+\mn}+g_c^2(\mn-\mpp)\bigg(\frac{1}{s-\mn^2}+\frac{1}{u-\mn^2}\bigg), \notag\\
 A^-&=g_c^2(\mn-\mpp)\bigg(\frac{1}{s-\mn^2}-\frac{1}{u-\mn^2}\bigg),\notag\\
 B^+&=-g_c^2\bigg(\frac{1}{s-\mn^2}-\frac{1}{u-\mn^2}\bigg), \notag\\
 B^-&=-\frac{2g_c^2}{(\mpp+\mn)^2}-g_c^2\bigg(\frac{1}{s-\mn^2}+\frac{1}{u-\mn^2}\bigg).
\end{align}
The constant terms in $A^+$ and $B^-$ generalize $g^2/\mN$ and $-g^2/2\mN^2$ in~\eqref{Born_terms}, they are not included in the definition of the partial-wave-projected
Born terms needed in~\eqref{MO}.
However, the implied shift in the subthreshold parameters is completely negligible, e.g., 
in the case of $B^-$ we find $\Delta b_{00}^-=-3\times 10^{-3}\mpi^{-2}$, to be compared with $b_{00}^-=10.49(11)\mpi^{-2}$~\cite{Hoferichter:2015hva}.
Using the projection~\eqref{tprojform}, with all kinematic factors understood to be defined by the charged-particle masses, 
we obtain for the partial-wave projection of the Born terms~\eqref{AB_isospin}
\begin{align}
 \tilde N_+^J(t)&=\frac{g_c^2}{4\pi}\mpp\bigg(\frac{Q_J(\tilde y)}{(p_tq_t)^J}\bigg(\tilde y+\frac{p_t}{q_t}\frac{\mn-\mpp}{\mpp}\bigg)-\delta_{J0}\bigg),\notag\\
 \tilde N_-^J(t)&=\frac{g_c^2}{4\pi}\frac{\sqrt{J(J+1)}}{2J+1}\frac{Q_{J-1}(\tilde y)-Q_{J+1}(\tilde y)}{(p_tq_t)^J},
\end{align}
with Legendre functions of the second kind
\beq
Q_J(z)=\frac{1}{2}\int\limits_{-1}^1\diff x\;\frac{P_J(x)}{z-x},
\eeq
and
\beq
\tilde y=\frac{t-2\mpi^2+2(\mn^2-\mpp^2)}{4p_tq_t}.
\eeq
These corrections are potentially relevant because in the vicinity of $\tpi$ they scale as $(\mn-\mpp)/\mpi\sim 1\%$ instead of $(\mn-\mpp)/\mpp\sim 0.1\%$.
However, near threshold, where the effect is most pronounced, the difference in the spectral functions
\begin{align}
 \Delta\Im G_E^v(t)&=\frac{q_t^3}{\mN\sqrt{t}}\big|F_\pi^V(t)\big|\Delta\tilde N_+^1(t)\cos\delta^1_1(t)\theta\big(t-\tpi\big),\notag\\
  \Delta\Im G_M^v(t)&=\frac{q_t^3}{\sqrt{2t}}\big|F_\pi^V(t)\big|\Delta\tilde N_-^1(t)\cos\delta^1_1(t)\theta\big(t-\tpi\big),
\end{align}
is strongly suppressed by phase space, and the same is true for the Born-term contribution to the dispersive integrals in~\eqref{MO}.
In the end, the remaining isospin-violating effect due to the proton--neutron mass difference in the Born terms is much smaller than the uncertainty from the subthreshold parameters and phase shifts, and can therefore be safely ignored.
These findings agree with the analysis of proton--neutron-mass-difference effects in the context of charge symmetry breaking in the nucleon form factors~\cite{Kubis:2006cy,Wagman:2014nfa}. 

\subsection{Subthreshold parameters}

The analysis of isospin violation in the low-energy parameters naturally proceeds in ChPT. In the context of the $\pi N$ scattering lengths and the $\pi N$ $\sigma$-term large effects
proportional to the pion mass difference $\Delta_\pi=\mpi^2-\mpii^2=2e^2\Fpi^2 Z$ have been observed~\cite{Gasser:2002am,Hoferichter:2009ez,Hoferichter:2015dsa}, due to an enhancement by $\pi$ and numerical prefactors over the naively expected chiral scaling. Such corrections cannot appear in the spectral function of the nucleon form factors at one-loop order, but at two-loop level 
neutral-pion loops are allowed. They lead to shifts in the subthreshold parameters in~\eqref{MO} proportional to $\Delta_\pi$, to be extracted
from~\cite{Hoferichter:2009gn} in the following. Throughout, we follow the nomenclature of~\cite{Gasser:2002am} for the low-energy constants, apart from the $\bar d_i$, for which we use the conventions of~\cite{Ecker:1995rk,Fettes:1998ud} 
more frequently employed in isospin-symmetric ChPT studies of $\pi N$ scattering.

As a first step, we need to identify the form of the Born terms in the presence of electromagnetic corrections to be subtracted from the $\pi N$ amplitudes in the definition of the subthreshold parameters. 
Starting from the Goldberger--Treiman discrepancy in the form
\beq
\label{GT_disc}
g=\frac{\mN \ga}{\Fpi}\bigg(1-\frac{2\mpii^2\bar d_{18}}{\ga}\bigg),
\eeq
we can work backwards to determine the explicit form of the Born terms to be subtracted in the chiral expansion. In particular, we need the renormalization
\begin{align}
 F_\pi&=F\bigg\{1+\frac{\mpii^2}{\Fpi^2}l_4^\text{r}+\frac{10}{9}e^2\big(k_1^\text{r}+k_2^\text{r}\big)+2e^2k_9^\text{r}\\
 &-\frac{\mpi^2}{32\pi^2\Fpi^2}\log\frac{\mpi^2}{\mu^2}-\frac{\mpii^2}{32\pi^2\Fpi^2}\log\frac{\mpii^2}{\mu^2}\bigg\},\notag\\
 \ga&=\tilde g\bigg\{1+\frac{4 d_{16}^\text{r}\mpii^2}{\ga}-\frac{\ga^2\mpi^2}{16\pi^2\Fpi^2}+\frac{e^2\Fpi^2}{\ga}\Big(g_1^\text{r}+g_2^\text{r}+\frac{g_{11}^\text{r}}{2}\Big)\notag\\
 &-\frac{\big(3\ga^2+1)\mpi^2}{32\pi^2\Fpi^2}\log\frac{\mpi^2}{\mu^2}-\frac{\big(\ga^2+1)\mpii^2}{32\pi^2\Fpi^2}\log\frac{\mpii^2}{\mu^2}\bigg\},\notag
\end{align}
with renormalized couplings depending on the renormalization scale $\mu$.
In analogy to $\bar d_{18}$ above, we also define 
\begin{align}
  d_i^\text{r}&=\bar d_i+\frac{\beta_i}{32\pi^2\Fpi^2}\log\frac{\mpi^2}{\mu^2},\notag\\
 g_i^\text{r}&=\bar g_i+\frac{\eta_i}{32\pi^2\Fpi^2}\log\frac{\mpi^2}{\mu^2},\notag\\
  k_i^\text{r}&=\bar k_i+\frac{\sigma_i}{32\pi^2}\log\frac{\mpi^2}{\mu^2},
\end{align}
with $\beta$-functions $\beta_i$, $\eta_i$, and $\sigma_i$,
and only the $Z$-dependent terms in the $\eta_i$ and $\sigma_i$ as listed in~\cite{Gasser:2002am} should be kept in order to isolate the $\Delta_\pi$ effects. 

In these conventions, one obtains a renormalized amplitude for the $\pi^\pm p$ channels at leading-loop order $\Order(p^3)$~\cite{Hoferichter:2009gn}, including $\Delta_\pi$ effects but no virtual photons.
In particular, we verified that for $e\to 0$ the known chiral expansion of the subthreshold parameters is reproduced. 
To proceed beyond the isospin limit we observe that all additional pole terms can be absorbed into a simple redefinition of the axial coupling
\beq
\label{shift}
\tilde g\to\tilde g+e^2\Fpi^2\Big(\frac{\bar g_{11}}{2}-\frac{2}{\Fpi^2}\ga\bar k_9\Big), 
\eeq
which amounts to a generalized Goldberger--Treiman discrepancy of
\beq
\label{GT_disc_IV}
g_c=\frac{\mN \ga}{\Fpi}\bigg(1-\frac{2\mpii^2\bar d_{18}}{\ga}-\frac{e^2\Fpi^2\bar g_{11}}{2\ga}+2e^2\bar k_9\bigg).
\eeq
Note that just as $\bar d_{18}$, the low-energy constants $\bar g_{11}$ and $\bar k_9$ do not involve chiral logarithms, i.e., the original $g_{11}$ and $k_9$ are finite.

After the shift~\eqref{shift} all pole terms disappear and the corrections to the subthreshold parameters relevant for the $P$-waves become
\begin{align}
\label{subthreshold_IV}
 \Delta a_{00}^-&=e^2\big(\bar g_6+\bar g_8\big)+\frac{e^2}{\Fpi^2}\bar k_9-\frac{4\Delta_\pi}{\Fpi^2}\bar d_5-\frac{\ga^4\Delta_\pi}{96\pi^2\Fpi^4},\notag\\
 \Delta a_{01}^-&=\frac{\ga^4\Delta_\pi}{192\pi^2\Fpi^4\mpi^2},
\end{align}
while $b_{00}^-$ and $b_{01}^-$ remain unaffected at $\Order(p^3)$.
With $\bar d_5=0.14\GeV^2$ from~\cite{Hoferichter:2015hva} and ignoring the $g_i$ and $k_i$ contributions, we obtain numerically
\beq
 \Delta a_{00}^-=-3\times 10^{-3}\mpi^{-2},\qquad \Delta a_{01}^-=0.5\times 10^{-3}\mpi^{-4},
\eeq
to be compared with the RS results~\cite{Hoferichter:2015hva}
\begin{align}
a_{00}^-+b_{00}^-&=1.411(15)\mpi^{-2},\qquad a_{00}^-=-9.08(12)\mpi^{-2},\notag\\
a_{01}^-+b_{01}^-&=-0.141(5)\mpi^{-4},\qquad a_{01}^-=-0.35(2)\mpi^{-4},
\end{align}
where the combinations $a_{0n}^-+b_{0n}^-$ become relevant in $f^1_+(t)$ for small values $t\ll \tN$.
We conclude that at the present level of accuracy the isospin-violating effects in the subthreshold parameters are too small to matter. 
In particular, enhancements by $\pi$ or numerical prefactors do not occur, so that the remaining shifts~\eqref{subthreshold_IV} become negligible compared to
the RS uncertainties in the subthreshold parameters.

\subsection{$\boldsymbol{\rho}$--$\boldsymbol{\omega}$ mixing}
\label{sec:rho_omega}

\begin{figure}[t!]
\centering
\includegraphics[width=0.35\linewidth]{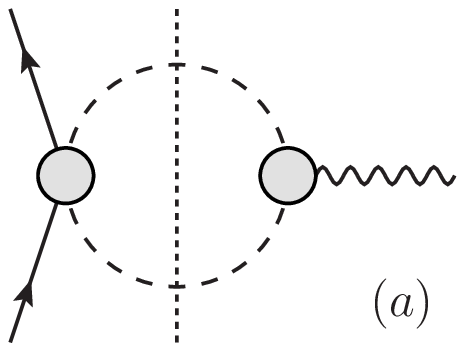}\qquad 
\includegraphics[width=0.35\linewidth]{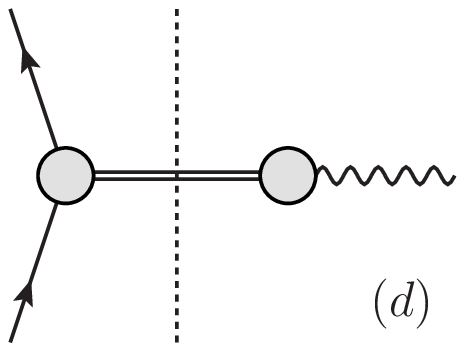}\\[0.2cm] 
\includegraphics[width=0.35\linewidth]{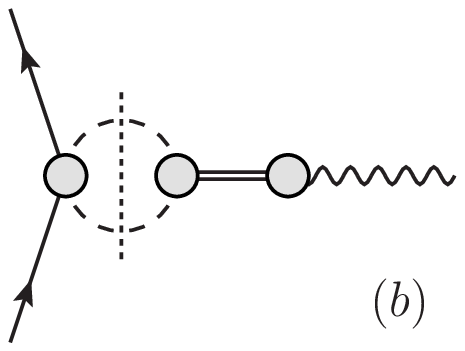}\qquad 
\includegraphics[width=0.35\linewidth]{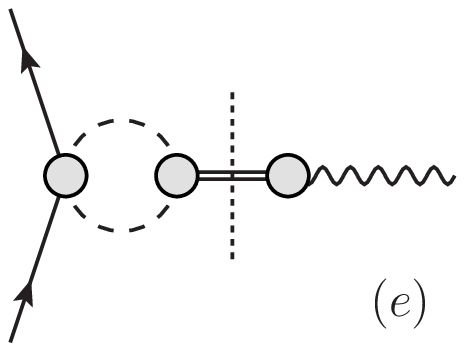}\\[0.2cm] 
\includegraphics[width=0.35\linewidth]{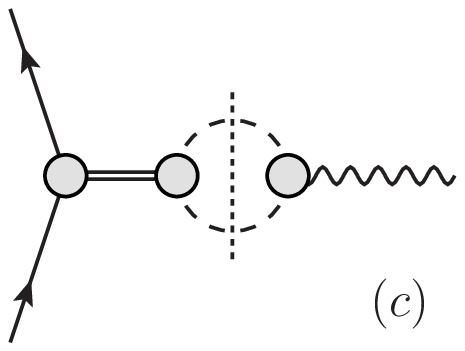}\qquad 
\includegraphics[width=0.35\linewidth]{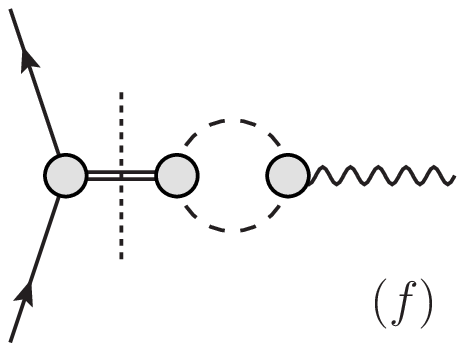} 
\caption{$(a)$: unitarity diagram for the $2\pi$ cut, $(d)$: unitarity diagram for the $3\pi$ cut approximated by $\omega$ exchange, $(b)$, $(c)$: isospin-violating contributions to the $2\pi$ cut due to the $\omega\to2\pi$ coupling, 
$(e)$, $(f)$: isospin-violating contributions to the $3\pi$ cut.
Double lines denote the $\omega$ propagator, otherwise notation as in Fig.~\ref{fig:unitarity}.}
\label{fig:unitarity_2pi_3pi}
\end{figure}

$\rho$--$\omega$ mixing occurs because the $\omega$ has a non-vanishing branching fraction to $2\pi$~\cite{Agashe:2014kda}
\beq
\text{BR}(\omega\to\pi^+\pi^-)=1.53^{+0.11}_{-0.13}\%,
\eeq
which leads to observable mixing effects in the pion vector form factor as measured in $e^+e^-$ scattering. In principle, this implies that the $2\pi$ and $3\pi$ channels are coupled and can no longer be studied separately, as reflected by the fact that the inclusion of the $\omega$ propagator in~\eqref{FpiV} spoils the cancellation of imaginary parts on the right-hand side of~\eqref{unitarity_relation}.
The origin of this behavior is illustrated in Fig.~\ref{fig:unitarity_2pi_3pi}: Eq.~\eqref{unitarity_relation} corresponds to the $2\pi$ cut in diagram $(a)$, while the $\omega$ admixture to the pion vector form factor is represented by diagram $(b)$. In the same way, there should be an isospin-violating $\omega$-exchange contribution to the $\pi\pi\to\bar N N$ amplitude, see diagram $(c)$. 
Approximating the $3\pi$ intermediate states by $\omega$ exchange, the primary $3\pi$ cut corresponds to putting the $\omega$ propagator on-shell, as in diagram $(d)$. In addition, there will be isospin-violating corrections represented by the exact same diagrams as in the left column, the sole difference being that the $\omega$ propagator is cut first. Only the combination of all diagrams will result in a consistent spectral function. 

However, we recall that in the presence of isospin violation isoscalar and isovector contributions are to be classified according to their nucleon couplings. In particular, the $\omega$ couplings to the nucleon do not distinguish between proton and neutron, so that, in these conventions, diagrams $(c)$ and $(f)$ should, together with diagram $(d)$, be included in the isoscalar spectral functions. 
By definition, the corresponding isospin-violating effects depend on the $\omega$ couplings to the nucleon~\cite{Kubis:2006cy,Wagman:2014nfa}.
As far as the isovector spectral functions are concerned, the only new contribution is given by diagram $(e)$, with a sub-amplitude $\bar N N\omega$ that can again be reconstructed by means of a dispersion relation. Taking everything together we find 
\begin{align}
\label{unitarity_final}
 \Im G_E^v(t)&=\frac{q_t^3}{\mN\sqrt{t}}|\Omega_1^1(t)||f_+^1(t)|\theta\big(t-\tpi\big)\notag\\
 &\qquad\times\bigg( 1+\alpha t + \frac{\eps\,t}{M_\omega^2+iM_\omega\Gamma_\omega-t} \bigg)\notag\\
&+\eps \,\Im\bigg(\frac{t}{M_\omega^2-i M_\omega  \Gamma_\omega-t}\bigg)\notag\\
&\qquad\times\frac{1}{\pi}\int\limits_{\tpi}^\infty \diff t'\frac{\frac{q_t'^3}{\mN\sqrt{t'}}|\Omega_1^1(t')||f_+^1(t')|}{t'-t-i\eps},\notag\\
 \Im G_M^v(t)&=\frac{q_t^3}{\sqrt{2t}}|\Omega_1^1(t)||f_-^1(t)| \theta\big(t-\tpi\big)\notag\\
 &\qquad\times \bigg( 1 +\alpha t+ \frac{\eps\,t}{M_\omega^2+iM_\omega\Gamma_\omega-t} \bigg)\notag\\
 &+\eps \,\Im\bigg(\frac{t}{M_\omega^2-i M_\omega  \Gamma_\omega-t}\bigg)\notag\\
 &\qquad\times\frac{1}{\pi}\int\limits_{\tpi}^\infty \diff t'\frac{\frac{q_t'^3}{\sqrt{2t'}}|\Omega_1^1(t')||f_-^1(t')|}{t'-t-i\eps}.
\end{align}
Strictly speaking the narrow-width approximation for the $\omega$ only works at $t=M_\omega^2$, i.e.
\beq
\Im\bigg(\frac{t}{M_\omega^2-i M_\omega  \Gamma_\omega-t}\bigg)\to \pi M_\omega^2\delta\big(t-M_\omega^2\big).
\eeq
Separating real and imaginary part of the dispersive integral in~\eqref{unitarity_final} and treating the $\omega$ propagator from the representation of the pion form factor in the same way
proves that at $t=M_\omega^2$ indeed the imaginary parts on the right-hand side cancel. Beyond $t=M_\omega^2$, the use of the same Breit--Wigner approximation in both cases ensures that this cancellation also works for arbitrary $t\geq \tpi$. Since the real part of the diagram-$(e)$ contribution cannot be used down to $\tpi$---it exhibits the wrong threshold behavior---we introduce a step function that puts this term to zero below $t=9\mpi^2$, the nominal threshold of the $3\pi$ channel. Given the small width of the $\omega$ the choice of this threshold is immaterial, all effects are localized closely around $t=M_\omega^2$. For the sum rules studied in Sect.~\ref{sec:sum_rules}, we also compared the outcome using a strict $\delta$-function or a Breit--Wigner representation with a finite width, with the result 
that the difference in the integrated quantity is hardly visible.

\section{Results for the spectral functions}
\label{sec:spectral}

\begin{figure}[t!]
\centering
\includegraphics[width=0.95\linewidth,clip]{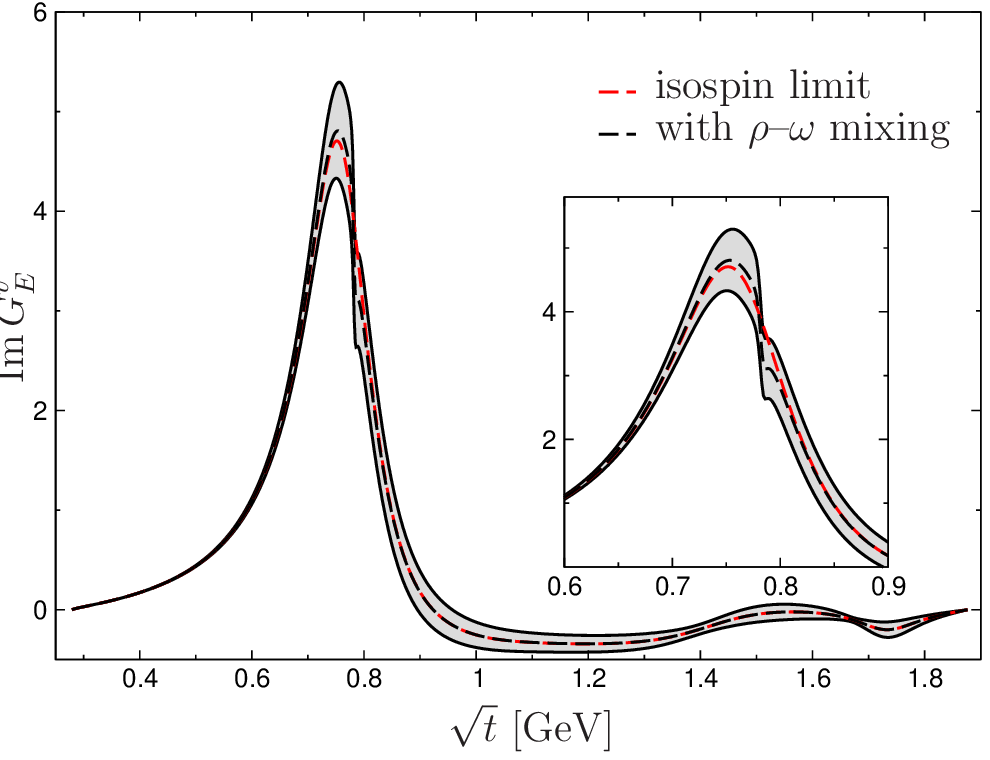}\\[0.2cm]
\includegraphics[width=0.95\linewidth,clip]{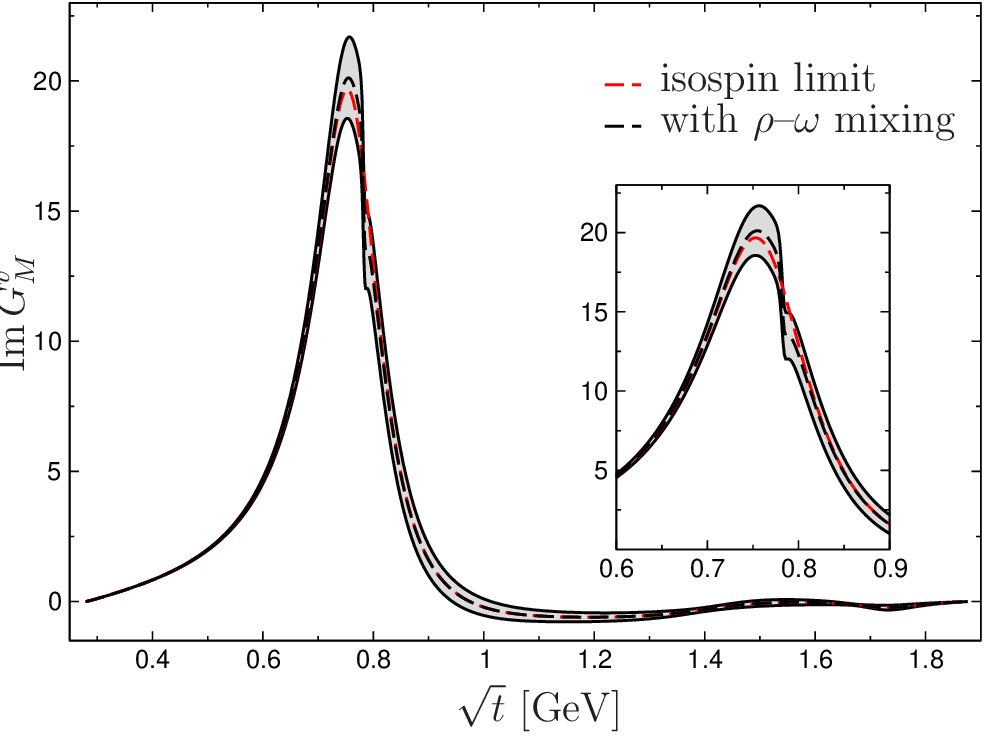}
\caption{Isovector spectral functions for $G_E^v$ and $G_M^v$. The black dashed line gives our central solution, the gray band the uncertainty estimate, and the red dashed line the result if $\rho$--$\omega$ mixing is turned off. The insets magnify the region around the $\rho$ peak. 
Text files of the results are available as supplementary material.}
\label{fig:ImG}
\end{figure}

Our final result for the isovector spectral functions, derived based on~\eqref{unitarity_final}, are shown in Fig.~\ref{fig:ImG}. The uncertainty band covers the following effects, added in quadrature:
\begin{enumerate}
 \item subthreshold parameters,
 \item $\pi\pi$ phase shift $\delta^1_1(t)$,
 \item data for pion vector form factor $F_\pi^V(t)$.
\end{enumerate}
The uncertainties from the RS determination of the subthreshold parameters, $b_{00}^-$, $b_{01}^-$, $a_{00}^-$, $a_{01}^-$, are propagated using their full covariance matrix. This is the dominant source of uncertainty below $1\GeV$. 
As detailed in Sect.~\ref{sec:formalism}, we use three variants of $\delta^1_1(t)$, consistent with the corresponding input in the RS solution, one of which includes the effects of $\rho'$, $\rho''$ in an elastic approximation and is taken as our central solution. $F_\pi^V(t)$ is fit to the
latest experimental results from BaBar~\cite{Aubert:2009ad}, KLOE~\cite{Babusci:2012rp}, and BESIII~\cite{Ablikim:2015orh}, restricting the fit region to $\sqrt{t}\leq 1\GeV$, where our representation can be rigorously justified. The central value is defined as the average between the three experiments, with uncertainties estimated by the variation among them. 
Although we do include some information on $4\pi$ intermediate states by means of the $\pi\pi$ phase shift, we stress that above $1\GeV$ we do not have a complete description. Here, the interplay with the $4\pi$ channel might introduce additional uncertainties.

Finally, Fig.~\ref{fig:ImG} also shows the results when $\rho$--$\omega$ mixing is switched off. The distortions due to this remaining isospin-violating effect are found to be relatively minor, but this is also a consequence of~\eqref{unitarity_final}: the new contribution from cutting the $\omega$ propagator tends to cancel $\rho$--$\omega$ mixing in the pion form factor.
In addition to using consistent input for the $\pi\pi$ phase shift in all parts of the calculation, updating the $\pi N$ partial waves to the latest phenomenological results, and providing thorough uncertainty estimates, the improved treatment of $\rho$--$\omega$ mixing constitutes a major advancement compared to previous analyses. 

Given that the final uncertainties are dominated by the
$\pi\pi\to\bar N N$ partial waves, the comparison to the KH80 amplitudes
as detailed in~\cite{Hoferichter:2015hva} implies that the resulting spectral
functions are consistent with~\cite{Hoehler:1983}
within uncertainties once the pion form factor is updated.
The slight enhancement of the spectral function near the $\rho$ peak
observed in~\cite{Belushkin:2005ds}
compared to our result is related to the assumption
$(F_\pi^V)^*f_\pm^1\to |F_\pi^V||f_\pm^1|$ in the unitarity relation,
and does not occur in the full treatment of $\rho$--$\omega$ mixing
according to~\eqref{unitarity_final}.

\section{Sum rules}
\label{sec:sum_rules}

\begin{table}
\centering
\renewcommand{\arraystretch}{1.3}
\begin{tabular}{crrr}
\toprule
 & $\Lambda=1\GeV$ & $\Lambda=2\mN$ & exact\\\midrule
$G_E^v(0)$ & $0.76(8)$ & $0.68(11)$ & $0.5$\\
$G_M^v(0)$ & $3.34(25)$ & $3.21(30)$ & $2.35$\\
$\langle r^2_E\rangle^v \ [\fm^2]$ & $0.418(32)$ & $0.405(36)$ &\\
$\langle r^2_M\rangle^v \ [\fm^2]$ & $1.83(10)$ & $1.81(11)$ &\\
\bottomrule
\end{tabular}
\caption{Sum rules for form-factor normalizations and radii, evaluated with integral cutoff $\Lambda$.}
\label{table:sum_rules}
\renewcommand{\arraystretch}{1.0}
\end{table}

Based on the generic form of the dispersion relations~\eqref{disp}, it is straightforward to derive sum rules for the normalization and derivatives of the isovector form factors. We obtain
\begin{align}
 G_E^v(0)&=\frac{1}{\pi}\int\limits_{\tpi}^\infty \diff t'\frac{\Im G_E^v(t')}{t'}=\frac{1}{2},\notag\\
 G_M^v(0)&=\frac{1}{\pi}\int\limits_{\tpi}^\infty \diff t'\frac{\Im G_M^v(t')}{t'}=\frac{1+\kappa_p-\kappa_n}{2},\notag\\
 \langle r^2_E\rangle^v&=\frac{6}{\pi}\int\limits_{\tpi}^\infty \diff t'\frac{\Im G_E^v(t')}{t'^2}=\frac{1}{2}\Big[\langle r^2_E\rangle^p-\langle r^2_E\rangle^n\Big],\notag\\
 \langle r^2_M\rangle^v&=\frac{6}{\pi}\int\limits_{\tpi}^\infty \diff t'\frac{\Im G_M^v(t')}{t'^2}\notag\\
 &=\frac{1}{2}\Big[(1+\kappa_p)\langle r^2_M\rangle^p-\kappa_n\langle r^2_M\rangle^n\Big],
\end{align}
where the magnetic moments in $\langle r^2_M\rangle^v$ compensate for the conventional normalization in~\eqref{radii}.
Note that the sum rules for the radii remain unchanged if a once-subtracted dispersion relation is used instead of the unsubtracted one in~\eqref{disp}. As consequence, 
the sum-rule results regarding the proton radius puzzle are independent of the dispersion relation assumed for the form factors.

Our results for these sum rules, using the $\pi\pi$ spectral functions presented in the previous section, are summarized in Table~\ref{table:sum_rules}. To estimate the sensitivity of the integrals to the high-energy tail, we show results both for an integral cutoff $\Lambda=1\GeV$ and $\Lambda=2\mN$. As expected, the sum rules for the normalization converge slowly at best, for $\Lambda=2\mN$ our central values differ from the exact results by about $30\%$. In contrast, the shift observed in the radii between $\Lambda=1\GeV$ and $\Lambda=2\mN$ is quite small, in fact, smaller than the uncertainty estimate from the spectral function below $1\GeV$. To make this statement more quantitative, we consider an additional effective narrow resonance with mass $M_R$ in the spectral function 
\beq
\Im G_{E/M}^v(t)=c^v_{E/M}\pi M_R^2\delta(t-M_R^2),
\eeq
with $c_E^v=-0.07$, $c_M^v=-0.55$ to make the sum rules for the normalization agree within $1\sigma$ with their expected values. The impact on the radii,
\beq
\Delta \langle r^2_{E/M}\rangle^v=\frac{6c_{E/M}^v}{M_R^2},
\eeq
can be made arbitrarily small by taking the resonance mass to infinity, but even in the most pessimistic scenario where $M_R=(1.4\ldots1.6)\GeV$ is varied within the energy range where $\rho'$, $\rho''$ become relevant we find $\Delta \langle r^2_E\rangle^v=-(0.006\ldots0.008)\fm^2$, $\Delta \langle r^2_M\rangle^v=-(0.05\ldots0.07)\fm^2$. 
In particular the electric radius as predicted from the sum rules should therefore be reasonably stable with respect to the contribution from higher intermediate states. 

First, we turn to the magnetic radius. Although~\cite{Bernauer:2013tpr} quotes a much lower value, $r_M^p=0.777(17)\fm$, the tensions between different analyses are in general less severe than for the electric radius, e.g., comparing $r_M^p=0.86^{+0.02}_{-0.03}\fm$, $r_M^n=0.88(5)\fm$~\cite{Lorenz:2012tm} and $r_M^p=0.87(2)\fm$, $r_M^n=0.89(3)\fm$~\cite{Epstein:2014zua}. 
The corresponding results for the isovector combination, $\langle r^2_M\rangle^v=1.78^{+0.10}_{-0.11}\fm^2$ and $\langle r^2_M\rangle^v=1.81(7)\fm^2$,\footnote{Due to isospin considerations, the errors for proton and neutron radii in~\cite{Epstein:2014zua} are not independent. Since we do not include this correlation, the error in the isovector combination should be taken as indicative.} are in good agreement with our sum-rule value 
$\langle r^2_M\rangle^v=1.81(11)\fm^2$.

In contrast to the proton radius, the electric charge radius of the neutron is far less contentious, the PDG quotes $\langle r_E^2\rangle^n=-0.1161(22)\fm^2$~\cite{Agashe:2014kda}, mainly based on~\cite{Koester:1995nx,Kopecky:1997rw}. With the neutron radius determined, the proton radius puzzle can be translated into an isovector radius puzzle, with $r_E^p=0.841\fm$ and $r_E^p=0.876\fm$ corresponding to $\langle r^2_E\rangle^v=0.412\fm^2$ and $\langle r^2_E\rangle^v=0.442\fm^2$, respectively.
Our sum-rule result $\langle r^2_E\rangle^v=0.405(36)\fm^2$ could thus be interpreted as a mild preference for the small radius, but in view of the uncertainty estimate the constraint from the $\pi\pi$ spectral function alone is clearly not sufficient to distinguish between the two scenarios.

\section{Conclusions}
\label{sec:conclusions}

In this paper we provided an updated analysis of the $\pi\pi$ contribution to the isovector spectral functions of the nucleon electromagnetic form factors, based on the $\pi\pi\to\bar N N$ partial waves as determined recently from Roy--Steiner equations and including the most recent experimental results for the pion vector form factor. Special attention is paid towards consistency of the various input quantities, in particular as regards the $\pi\pi$ phase shift used in the calculation, and towards estimating the potential impact of isospin-violating corrections. 
The constraints provided by our results for the spectral functions, which, for the first time, include a thorough uncertainty estimate, should prove valuable in future analyses of the nucleon form factors. 

As an application, we studied the $\pi\pi$ saturation of the sum rules for form-factor normalizations and radii. While, as expected, the sum rules for the normalizations are at best slowly convergent, those for the radii prove to be more stable, with a resulting value for the isovector magnetic radius in good agreement with previous determinations. 
Taking the neutron electric radius from the literature, we find a slight preference for a small proton charge radius, but the uncertainties in the spectral-function constraint alone are too large to draw firm conclusions. However, the strategy of concentrating on the isovector radius offers synergies with lattice calculations, where due to disconnected diagrams in the isoscalar form factor the isovector combination can be determined more accurately. With input from phenomenology for the neutron radius, the isovector part alone already has important implications for the interpretation of the proton radius puzzle.

\section*{Acknowledgements}

Financial support by
the DFG (SFB/TR 16, ``Subnuclear Structure of Matter,''
SFB/TR 110, ``Symmetries and the Emergence of Structure in QCD,''
SFB 1245, ``Nuclei: From Fundamental Interactions to Structure and Stars''), 
and the DOE (Grant No.\ DE-FG02-00ER41132) 
is gratefully acknowledged.
The work of UGM was supported in part by The Chinese Academy of Sciences 
(CAS) President's International Fellowship Initiative (PIFI) grant no.\ 2015VMA076.
\esp

\appendix

\end{document}